\documentstyle[12pt]{article}
\newcommand{\h}{\hspace*{5 ex}}
\newcommand{\disp}{\displaystyle}
\newcommand{\bi}{\bibitem}
\newcommand{\vs}{\vspace*}

\newcommand{\drm}{{\rm d}}
\newcommand{\ove}{\overline}
\newcommand{\pa}{\partial}

\begin{document}

\centerline{\bf MORE ABOUT TUNNELLING TIMES, THE DWELL TIME,}
\vspace*{2mm}
\centerline{\bf AND THE ``HARTMAN EFFECT"$^{(*)}$}

\footnotetext{$^{(*)}$ Work partially supported by
INFN, MURST and CNR (Italy), by CNPq (Brazil), and by
the I.N.R. (Kiev, Ukraine).}

\vspace*{0.9 cm}

\centerline{Vladislav S. OLKHOVSKY}

\vspace*{0.2 cm}

{\small
\centerline{{\em Institute for Nuclear Research, Ukrainian Academy of Sciences,
Kiev, Ukraine;}}
\centerline{{\em and I.N.F.N., Sezione di Catania, 57 Corsitalia, Catania,
Italy.}}
}

\vspace*{0.4 cm}

\centerline{Erasmo RECAMI}

\vspace*{0.2 cm}

{\small
\centerline{{\em Facolt\`{a} di Ingegneria, Universit\`{a} Statale di
Bergamo, Dalmine (BG), Italy;}}
\centerline{{\em I.N.F.N., Sezione di Milano, Milan, Italy; \ and}}
\centerline{{\em Dept.$\!$ of Applied Mathematics, State University at
Campinas, Campinas, S.P., Brazil.}}
}

\vspace*{0.3 cm}

\centerline{Fabio RACITI}

\vspace*{0.2 cm}

{\small
\centerline{\em Dipartim. di Fisica, Universit\`a di Catania, Catania, Italy.}

\vspace*{.3 cm}

\centerline{ and }

\vspace*{.3 cm}

\centerline{Aleksandr K. ZAICHENKO}

\vspace*{0.2 cm}

{\small
\centerline{{\em Institute for Nuclear Research, Ukrainian Academy of Sciences,
Kiev, Ukraine.}}

\vspace*{1.2 cm}

{\bf Abstract  \  --} In a recent review paper [{\em Phys. Reports} {\bf 214}
(1992) 339] we proposed, within conventional quantum mechanics, new
definitions for the sub-barrier tunnelling and reflection times. \ Aims of the
present paper are: \ (i) presenting and analysing
the results of various numerical calculations (based on our equations)
on the penetration and return times $<\tau_{\, \rm Pen}>$,
$<\tau_{\, \rm Ret}>$,
during tunnelling {\em inside} a
rectangular potential barrier, for various penetration depths $x_{\rm f}$; \
(ii) putting forth and discussing
suitable definitions, besides of the mean values, also of the {\em variances}
(or dispersions) ${\rm D} \, {\tau_{\rm T}}$ and ${\rm D} \,
{\tau_{\, \rm R}}$
for the time  durations of transmission and reflection processes; \
(iii) mentioning, moreover, that our definition $<\tau_{\rm T}>$ for
the average transmission time results to constitute an {\em improvement} of
the ordinary dwell--time ${\ove \tau}^{\rm Dw}$ formula: \ (iv) commenting,
at last, on the basis of our {\em new} numerical results,
upon some recent criticism by C.R.Leavens.  \ \ We stress that our numerical
evaluations {\em confirm} that our approach implied, and implies, the
existence of the
{\em Hartman effect}: an effect that in these days (due to the theoretical
connections between tunnelling and evanescent--wave propagation) is receiving
---at Cologne,
Berkeley, Florence and Vienna--- indirect, but quite interesting, experimental
verifications. \ Eventually, we briefly  analyze some other definitions of
tunnelling times.\hfill\break

PACS nos.: \ 73.40.Gk ; \ 03.80.+r ; \ 03.65.Bz .

\newpage

{\bf 1. -- Introduction}

\vs{0.5 cm}

\h In our review article$^{[1]}$
[{\em Phys.Rep.} {\bf 214} (1992) 339] we put forth an analysis of the main
theoretical definitions of the sub-barrier tunnelling and reflection times,
and proposed new definitions for such durations which seem to be
self-consistent within {\em conventional} quantum mechanics.$^{\# 1}$
\footnotetext{$^{\# 1}$ Let us take advantage of the present opportunity for
pointing out that a misprint entered our eq.(10) in ref.[1], whose last
term $ka$ ought to be eliminated. \ Moreover, due to an {\em editorial} error,
in
the footnote at page 32 of our ref.[16] the dependence of $G$ on $\Delta k$
disappeared, whilst in that paper we had assumed \ $G(k-\ove{k}) \equiv C \,
\exp [-(k-\ove{k})^2 / (\Delta k)^2]$.}
 \ In particular, the ``prediction" by our
theory$^{[1]}$ of the reality of the {\em Hartman effect\/}$^{[2]}$ in
tunnelling
processes has recently received ---due to the analogy$^{[3]}$ between
tunnelling electrons and evanescent waves--- quite interesting, even if
indirect,
experimental verifications at Cologne,$^{[4]}$ Berkeley,$^{[5]}$
Florence$^{[6]}$ and Vienna.$^{[6]}$

\h Main aims of the present paper are: \ presenting and analysing
the results of several
numerical calculations of the penetration and return times {\em inside} a
rectangular potential barrier during tunnelling (Sect.{\bf 3}); \ and
proposing new suitable formulae for the
distribution variances of the transmission and reflection times
(Sect.{\bf 2}).
\h The results of our numerical evaluations
seem to confirm that our approach is
physically acceptable, and that it moreover
implied, and implies, the existence of the so-called
{\em ``Hartman effect\/"} even for {\em non}--quasi-monochromatic packets.\\

\h The present research field,
however, is developing so rapidly and in such a controversial manner, that
it may be convenient to add here ---before all---  some brief
comments about a few papers appeared during the last three years:\\

\h (i) First, let us mention that we had overlooked a new expression
for the dwell--time ${\ove \tau}^{\rm Dw}$ derived by Jaworsky and
Wardlaw$^{[7,8]}$

\

\hfill{$
{\ove \tau}^{\rm Dw}(x_{\rm i},x_{\rm f};k) \; = \; \left(
{\int_{-\infty}^{\infty} {\rm d}t \, t \, J(x_{\rm f},t) \: - \:
\int_{-\infty}^{\infty} {\rm d}t \, t \, J(x_{\rm i},t)}\right)
\left({\int_{-\infty}^{\infty} {\rm d}t \; J_{\rm in}(x_{\rm i},t)}
\right)^{-1} \;\; ,$
\hfill} (1)

\

which is indeed equivalent$^{[7]}$ to our eq.(16) of ref.[1] (all
notations being defined therein):

\

\hfill{$
{\ove \tau}^{\rm Dw}(x_{\rm i},x_{\rm f};k) \; = \; \left(
{\int_{-\infty}^{\infty} {\rm d}t \int_{x_{\rm i}}^{x_{\rm f}} {\rm d}x \;
\rho (x,t)}\right) \left({\int_{-\infty}^{\infty} {\rm d}t \;
J_{\rm in}(x_{\rm i},t)} \right)^{-1} \;\; .$
\hfill} (2)

\

This equivalence
{\em reduces} the difference, between our definition
$<\tau_{\rm T}>$ of the
average transmission time ---under our assumptions---  and quantity
${\ove \tau}^{\rm Dw}$, {\em to} the difference
between the average made by using the positive--definite probability
density ${\rm d}t \, J_{+}(x,t)/\int_{-\infty}^{\infty}
{\rm d}t \, J_{+}(x,t)$
and the average made by using the ordinary ``probability density"
${\rm d}t \, J(x,t)/\int_{-\infty}^{\infty} {\rm d}t \, J(x,t)$. \ Generally
speaking, the last expression is {\em not} always positive definite,
as it was explained at page 350 of ref.[1], and hence does not possess
any direct physical meaning.\\

\h (ii) In ref.[9] an attempt was made to analyze the evolution of the
wave packet mean position $<x(t)>$ (``center of gravity"), averaged over
$\rho {\rm d}x$, during its tunnelling through a potential barrier. \ Let
us here observe that the conclusion to be found therein, about the absence
of a causal relation between the incident space centroid and its
transmitted equivalent, holds {\em only} when the contribution
coming from the barrier region to the space integral is negligible.\\

\h (iii) Let us also add that in ref.[10] it was analyzed the
{\em distribution}
of the transmission time $\tau_{\rm T}$ in a rather sophisticated way, which
is very similar to the dwell--time approach, however with an
{\em artificial}, abrupt switching on of the initial wave packet. \ \
We are going to propose, on the contrary, and in analogy with our
eqs.(30)-(31) in
ref.[1], the following expressions, as physically adequate definitions
for the {\em variances} (or dispersions)
${\rm D} \, \tau_{\rm T}$ and ${\rm D} \, \tau_{\, \rm R}$ of the transmission
and
reflection time [see Sect.{\bf 2}], respectively:

\

\hfill{${\rm D} \, \tau_{\rm T} \; \equiv \; {\rm D} \, t_{+}(x_{\rm f}) +
{\rm D} \, t_{+}(x_{\rm i})$\hfill} (3)

\

and

\

\hfill{${\rm D} \, \tau_{\, \rm R} \; \equiv \; {\rm D} \, t_{-}(x_{\rm i}) +
{\rm D} \, t_{-}(x_{\rm i}) \; ,$\hfill} (4)

where

\vskip 1. mm

\

\hfill{$
{\rm D} \, t_{\pm}(x) \; \equiv \; \disp{
\frac{\int_{-\infty}^{\infty} {\rm d}t \; t^2 \, J_{\pm}(x,t)}
{\int_{-\infty}^{\infty} {\rm d}t \, J_{\pm}(x,t)} } \; - \;
\disp{ \left( \frac{\int_{-\infty}^{\infty} {\rm d}t \; t \; J_{\pm}(x,t)}
{\int_{-\infty}^{\infty} {\rm d}t \, J_{\pm}(x,t)} \right) ^2} \; .$
\hfill} (5)

\

\vskip 1.5 mm

Equations (3)-(5) are based on the formalism expounded in ref.[11],
as well as on our definitions for $J_{\pm}(x,t)$ in ref.[1]. \ Of course, we
are supposing that the integrations over \ $J_{+}(x_{\rm f})  \, {\rm d}t$, \
$J_{+}(x_{\rm i})  \, {\rm d}t$ \ and \ $J_{-}(x_{\rm i})  \, {\rm d}t$
are independent of one another. \ We shall devote Sect.{\bf 2}, below, to
these problems, i.e., to the problem of suitably defining mean
values and variances  of durations, for various transmission and reflection
processes during tunnelling.\\

\h (iv) Below, in Sect.{\bf 4}, we shall briefly
re-analyse some other definitions of tunnelling durations.\\

\h Before going on, let us recall that several reasons ``justify" the
existence of different approaches to the definition of tunnelling times: \ \
(a) the problem of defining tunnelling durations is  closely connected
with that of defining a time operator, i.e., of introducing {\em time} as a
(non-selfadjoint) quantum mechanical observable, and subsequently of
adopting a general definition for collision durations in
 quantum mechanics. Such preliminary problems did receive some
 clarification in recent times (see, for
example, ref.[1] and citations [8] and [22] therein); \ \ (b) the motion of a
particle
tunnelling inside a potential barrier
is a purely quantum phenomenon, devoid of any classical, intuitive limit; \ \
(c) the various theoretical approaches may differ in the
choice of the boundary conditions or in the modelling of the experimental
situations.

\vfill
\newpage

{\bf 2. -- Mean values and Variances for various Penetration
and Return Times during tunnelling}

\vs{.5 cm}

\h In our previous papers,  we
proposed for the {\em transmission} and {\em reflection} times some
formulae which
imply ---as functions of the penetration depth--- integrations over time of
$J_{+}(x,t)$ and $J_{-}(x,t)$, respectively. \ Let us recall that
the total flux $J(x,t)$ inside a barrier consists of two components,
$J_{+}$ and $J_{-}$, associated with motion along the positive and the
negative $x$-direction, respectively. \ Work in similar directions
did recently appear in ref.[12].

\h Let us refer ourselves ---here--- to
tunnelling  and reflection processes of a particle by a potential barrier,
confining ourselves to one space dimension. \ Namely, let us study
the evolution of a wave packet
$\Psi(x,t)$,  starting from the initial state $\Psi_{\rm {in}}(x,t)$;
and follow the notation
introduced in ref.[1]. \ In the case of uni-directional  motions it is
already known$^{[13]}$ that the
flux density \ $J(x,t) \equiv {\rm Re}[ (i \hbar / m) \,
\Psi(x,t) \;
{\partial \Psi^{*}(x,t) / \partial x}]$ \ can be actually interpreted as the
probability that the
particle (wave packet) passes through position $x$ during a unitary
time--interval
centered at $t$, {\em as it easily follows from the continuity equation and
from the fact that quantity} \ $\rho(x,t) \equiv |\Psi(x,t)|^{2}$ \ {\em is
the probability density} for our
``particle" to be located, at time t, inside a unitary
space--interval centered at $x$. \ \ Thus, in order to determine
the {\em mean} instant at which a moving wave packet
$\Psi(x,t)$ passes through position $x$, we have to take the average
of the time variable $t$ with respect to the weight \ $w(x,t) \equiv
J(x,t)/ \int_{-\infty}^{\infty} J(x,t) \, \drm t$.

\h However, if the motion
direction can {\em vary}, then quantity $w(x,t)$ is no longer positive
definite, and moreover is not bounded, because of the variability of
the $J(x,t)$ sign.
 \ In such a case, one can introduce the two weights:

$$w_{+}(x,t) = J_{+}(x,t) \; \left[ \int_{-\infty}^{\infty} J_{+}(x,t)
 \: \drm t  \right]^{-1}
\eqno(6)$$

\

$$w_{-}(x,t) = J_{-}(x,t) \; \left[ \int_{-\infty}^{\infty} J_{-}(x,t)
 \: \drm t \right]^{-1}  \; ,
\eqno(7)$$

\

where $J_{+}(x,t)$ and $J_{-}(x,t)$ represent the positive and negative
parts of $J(x,t)$, respectively, which are bounded, positive--definite
functions, normalized to 1.  \
Let us show that, from the ordinary probabilistic interpretation of
$\rho (x,t)$ and from the well-known continuity equation

$$ {\partial{\rho(x,t)} \over {\partial t}} + {\pa J(x,t) \over \pa x} = 0
\ ,\eqno(8)$$

it  follows {\em also in this (more general) case} that quantities
$w_+$ and $w_-$,
represented by eqs.(6), (7), can be regarded as the probabilities that
our ``particle" passes through position $x$ during
a unit time--interval centered
at $t$ (in the case of forward and backward motion, respectively).

\h Actually, for those time intervals for which $J =
J_{+}$ or $J = J_{-}$, one can rewrite eq.(8) as follows:

$${{\partial {\rho_{>}(x,t)} \over \partial t} =
- {\pa J_{+}(x,t) \over {\pa x}}} \eqno(9.a)$$

\

$${{\partial {\rho_{<}(x,t)} \over \partial t} = -{\pa J_{-}(x,t) \over
\pa x} \ ,}
\eqno(9.b)$$

\

respectively. \ Relations (9.a) and (9.b) can be considered  as formal
definitions of \ $\partial {\rho_{>}} / \partial t$ and \ $\partial
{\rho_{<}} / \partial {t}$. \ \ Let us now integrate eqs.(9.a), (9.b)
over time from $-\infty$ to $t$; \ we obtain:

$$\rho_{>}(x,t)= -\int_{-\infty}^{t}
{\pa J_{+}(x,{t}') \over \pa x} \: \drm {t}' \eqno(10.a)$$

$$\rho_{<}(x,t)= -\int_{-\infty}^{t}
 {\pa J_{-}(x,{t}') \over \pa x} \: \drm t' \eqno(10.b)$$

\

with the initial conditions \ $\rho_{>}(x,-\infty)=\rho_{<}(x,-\infty)=0$.
 \ Then, let us introduce the quantities

$$N_{>} (x,\infty;t) \equiv \int_{x}^{\infty} \rho_{>}({x}',t) \, \drm {x}'
= \int_{-\infty}^{t} J_{+}(x,{t'}) \, \drm {t}' \ >0 \eqno(11.a)$$

\

$$N_{<} (-\infty,x;t) \equiv \int_{-\infty}^{x} \rho_{<}({x}',t) \,
\drm {x}' =
-\int_{-\infty}^{t} J_{-}(x,{t}') \, \drm {t}' \ >0 \ , \eqno(11.b)$$

\

which have  the meaning of probabilities for our ``particle" to be located
at time $t$ on the semi-axis  $(x,\infty)$ or $(-\infty,x)$
respectively, as functions of
the flux densities $J_{+}(x,t)$ or  $J_{-}(x,t)$,
provided that the normalization condition \
$\int_{-\infty}^{\infty}\rho(x,t) \drm x = 1$ \ is fulfilled.  \
The r.h.s.'s of eqs.(11.a) and (11.b) have been obtained by integrating the
r.h.s.'s of eqs.(10.a) and (10.b) and adopting the boundary conditions  \
$J_{+}(-\infty,t) = J_{-}(-\infty,t) = 0$. \  Now, by
differenciating eqs.(11.a) and (11.b) with respect to $t$, one obtains:

\
$${{\partial{N_{>}}(x,\infty,t) \over \partial{t}} =
J_{+}(x,t)  > 0 } \eqno(12.a)$$

\

$${{\partial{N_{<}}(x,-\infty,t) \over \partial{t}} =
- \, J_{-}(x,t)  > 0 } \ . \eqno(12.b)$$

{\em Finally}, from eqs.(11.a), (11.b), (12.a) and (12.b), one can infer that:

$${w_{+}(x,t) ={{\partial {N_{>}}(x,\infty;t)/\partial {t}
\over {N_{>}(x,-\infty,\infty)}}}} \eqno(13.a)$$

$${w_{-}(x,t) ={{\partial {N_{<}}(x,-\infty;t)/\partial {t}
\over {N_{<}(-\infty,x;\infty)}}}} \ , \eqno(13.b)$$

which justify the abovementioned probabilistic interpretation of $w_{+}(x,t)$
and $w_{-}(x,t)$. \ Let us notice, incidentally, that our approach does
{\em not} assume any ad hoc postulate, contrarily to what believed by the
author of ref.[14].\\

\h At this point, we can eventually define the {\em mean value} of the time
at which our ``particle"  passes
through position $x$,  travelling in the positive or negative direction of the
$x$ axis, respectively, as:

\
$$<t_{+}(x)> \ \equiv \ {{\int_{-\infty}^{\infty} t \, J_{+}(x,t) \,
\drm t  \over \int_{-\infty}^{\infty} J_{+}(x,t) \, \drm t }} \eqno(14.a)$$

\
$$<t_{-}(x)> \ \equiv \ {{\int_{-\infty}^{\infty} \, t J_{-}(x,t) \,
\drm t  \over \int_{-\infty}^{\infty} J_{-}(x,t) \, \drm t }} \eqno(14.b)$$

and, moreover, the {\em variances} of the distributions of these times as:

$${\rm D} \, t_+(x) \ \equiv \ {{\int_{-\infty}^{\infty} t^{2} J_{+}(x,t) \drm
t  \over
\int_{-\infty}^{\infty} J_{+}(x,t) \drm t }} - [<t_+(x)>]^{2} \eqno(15.a)$$

\

$${\rm D} \, t_-(x) \  \equiv \ {{\int_{-\infty}^{\infty} t^{2} J_{-}(x,t) \drm
t  \over
\int_{-\infty}^{\infty} J_{-}(x,t) \drm t }} - [<t_-(x)>]^{2} \ ,
\eqno(15.b)$$

in accordance with the proposal presented in refs.[1,15].

\h Thus, we have a formalism for defining mean values, variances
(and other central moments) related to the duration {\em distributions}
of all possible processes for a particle, tunnelling
through a potential barrier located in the interval $(0,a)$
along the $x$ axis; and not only for tunnelling, but also for all possible
kinds of collisions, with arbitrary energies and potentials.  \
For instance, we have that

$$<\tau_{\rm T}(x_{\rm i},x_{\rm f})> \ \equiv \ <t_+(x_{\rm f})> -
<t_+(x_{\rm i})> \eqno(16)$$

 \

with $-\infty < x_{\rm i} < 0$ \ and \ $a < x_{\rm f} < \infty$; \ and
therefore \ (as anticipated in eq.(3)) \ that

$${\rm D} \;
\tau_{\rm T}(x_{\rm i},x_{\rm f}) \; \equiv \; {\rm D} \, t_{+}(x_{\rm f}) +
{\rm D} \, t_{+}(x_{\rm i}) \ ,$$

for {\em transmissions} from
region $(-\infty,0)$ to region $(a,\infty)$ which we called$^{[1]}$ regions I
and III, respectively. \ \ Analogously, for the pure (complete) tunnelling
process one has:

 \
$$<\tau_{\rm Tun}(0,a)> \ \equiv \ <t_+(a)> - <t_+(0)> \eqno(17)$$

and

$${\rm D} \; \tau_{\rm Tun}(0,a) \; \equiv \; {\rm D} \, t_{+}(a) + {\rm D} \,
t_{+}(0) \ ; \eqno(18)$$

\

while one has

$$<\tau_{\, \rm Pen} (0,x_{\rm f})> \ \equiv \ <t_+(x_{\rm f})> - <t_+(0)>
\eqno(19)$$

\

and

$${\rm D} \; \tau_{\, \rm Pen} (0,x_{\rm f}) \ \equiv \
{\rm D} \, t_+(x_{\rm f}) + {\rm D} \, t_+(0)
\eqno(20)$$

\

[with \ $0<x_{\rm f}<a$] \ for {\em penetration} inside the barrier region
(which we called region II).  \ \ Moreover:

$$<\tau_{\, \rm Ret}(x,x)> \ \equiv \ <t_-(x)> -<t_+(x)> \eqno(21)$$

$${\rm D} \; \tau_{\, \rm Ret}(x,x) \ \equiv \ {\rm D} \, t_-(x)
 + {\rm D} \, t_+(x) \eqno(22)$$

[with \ $0<x<a$] \ for ``{\em return processes\/}"  inside the barrier.  \ At
last, for {\em reflections} in region I, we have that:

$$<\tau_{\, \rm R}(x_{\rm i},x_{\rm i})> \ \equiv \  <t_-(x_{\rm i})> -
<t_+(x_{\rm i})> \eqno(23)$$

[with $-\infty<x_{\rm i}<a$], \ and \ (as anticipated in eq.(4)) \ that \
${\rm D} \; \tau_{\, \rm R}(x_{\rm i},x_{\rm i}) \ \equiv \
{\rm D} \, t_-(x_{\rm i}) + {\rm D} \, t_+(x_{\rm i})$.

\h Let us stress that our definitions hold within
the framework of conventional quantum mechanics, without the introduction
of any new postulates, and with the single measure expressed by weights
$(13.a), \ (13.b)$ for all time averages (both in the initial and in the
final conditions).

\h According to our definition, the tunnelling phase time
(or, rather, the transmission duration), defined by the stationary phase
approximation for quasi-monochromatic particles, is meaningful {\em only}
in the limit $x_{\rm i} \rightarrow \infty$ when $J_{+}(x,t)$ is the flux
density
of the initial packet $J_{\rm in}$ of {\em incoming} waves  ({\em in absence
of reflected waves}, therefore).

\h Analogously, the dwell time, which can be represented (cf. eqs.(1),(2)) by
the
expression$^{[7,8,16]}$

$${{{\ove \tau}^{\rm Dw}(x_{\rm i},x_{\rm f}) = \left[
\int_{-\infty}^{\infty} t \; J(x_{\rm f},t) \; \drm t -
\int_{-\infty}^{\infty} t \; J(x_{\rm i},t) \; \drm t \; \right] \; \left[
\int_{-\infty}^{\infty} J_{\rm in}(x_{\rm i},t) \; \drm t \,
\right]^{-1} }} \ ,$$

\

with $-\infty < x_{\rm i} < 0$, and $x_{\rm f}>a$, \ is not
acceptable, generally speaking.  In fact,
the weight in the time averages is meaningful, positive definite
and normalized to 1 {\em only} in the rare cases when
$x_{\rm i} \longrightarrow-\infty$ and $J_{\rm in} = J_{\rm III}$ \ (i.e.,
when the barrier is transparent).

\vspace*{1. cm}

{\bf 3. -- Penetration and Return process durations, inside a rectangular
barrier, for tunnelling gaussian wave packets: Numerical results}

\vspace*{0.5 cm}

\h We put forth here the results of our calculations of mean durations for
various penetration (and return) processes, {\em inside} a rectangular
barrier, for tunnelling gaussian wave packets;
one of our aims being to investigate the
{\em tunnelling speeds}. \ In our calculations, the initial wave packet is

$$\Psi_{\rm in}(x,t) = \int_{0}^{\infty}G(k-\overline k) \;
\exp[ikx-iEt/\hbar] \; \drm k \eqno(24)$$

\

with

$$G(k- \overline k) \equiv C \exp [{- (k-\overline k)^2} /
{(2 \, \Delta k)^{2}}] \ , \eqno(25)$$

{\em exactly} as in ref.[8]; \ and with \
$E = \hbar^{2} k^{2}/2m$; \ quantity
$C$ being the normalization constant, and $m$ the electron mass. \
Our procedure  of integration was described
in ref.[16].

\

\h Let us express the penetration depth in {\aa}ngstroms, and
the penetration time in seconds. \ \ In Fig.1  we show the plots
corresponding to \ $a = 5 \;$\AA, \ for \ $\Delta k = 0.02$ and
$0.01 \; {\rm {\AA}}^{-1}$, \ respectively. \ The penetration time
$<\tau_{\, \rm Pen}>$ always tends to a {\em saturation} value.\\
\h In Fig.2 we show,
for the case \ $\Delta k = 0.01 \; {\rm {\AA}}^{-1}$, \ the plot corresponding
to \ $a = 10 \;$\AA. \ It is interesting that $<\tau_{\, \rm Pen}>$ is
practically {\em the same} (for the same $\Delta k$) for \ $a = 5$ and
$a = 10 \;$\AA, a result that confirm, let us repeat, the existence$^{[1]}$
of the so-called
{\em Hartman  effect\/}.$^{[2]}$ \ \ Let us add that, \ when varying the
parameter $\Delta k$ between $0.005$
and $0.15 \; {\rm \AA}^{-1}$ and letting $a$ to assume values even larger than
$10 \; {\rm \AA}$, \ analogous results have been always gotten. \ Similar
calculations have been performed (with
quite reasonable results) also for various energies $\ove{E}$ in the
range $1$ to $10$ eV.~$^{\# 2}$
\footnotetext{$^{\# 2}$ For the interested reader, let us recall that, when
integrating over d$t$, we used the interval \ $- 10^{-13} \:$s \ to \
$+10^{-13} \:$s \ (symmetrical
with respect to $t = 0$), \ very much larger than the temporal wave packet
extension. [Recall that the extension in time of a wave packets is of the
order of \ $1/(\ove{v} \,
{\Delta k}) = ({\Delta k} \, \sqrt{2 \ove{E} / m})^{-1} \simeq 10^{-16} \:$s].
\ Our ``centroid" has been always $t_{0} = 0; \; x_{0} = 0$. \ \  For
clarity's
sake, let us underline again that in our approach the initial
wave packet $\Psi_{\rm in} (x,t)$ is not regarded as prepared at a certain
instant of time, but it is expected to flow through any (initial) point
$x_{\rm i}$ during the infinite time interval ($-\infty, \;\; +\infty$),
even if with a {\em finite} time--centroid $t_0$. \ The value of such
centroid $t_0$ is essentially defined by the phase of the weight amplitude
$G(k- \ove{k})$, and in our case is equal to 0 when $G(k- \ove{k})$ is
real.}

\h In Figs.3, 4 and 5 we show the behaviour of the mean penetration
and return durations as function of the penetration depth
(with $x_{\rm i} = 0$ and
$0\le \, x_{\rm f} \equiv x \, \le a$), \ for barriers with height $V_{0} = 10$
eV
and width $a= 5 \;${\AA} or $10 \;$\AA. \ \
In Fig.3 we present the plots of
$<\tau_{\, \rm Pen}(0,x)>$ corresponding to different values of the mean
kinetic energy: \  $\overline E$ = 2.5 eV, $\;$5 eV and 7.5 eV (plots 1, 2
and 3, respectively) with $ \Delta k=0.02 {\rm \AA}$; \  and
$\overline E = 5$ eV with $\Delta k =0.04 {\rm \AA}^{-1}$ (plot 4),
always with $a= 5 {\rm \AA}$. \ \
In Fig.4 we show the plots of $<\tau_{\, \rm Pen}(0,x)>$,
corresponding to $a=5 {\rm \AA}$, with $\Delta k = 0.02 {\rm \AA}^{-1}$
and 0.04 ${\rm \AA}^{-1}$
(plots 1 and 2, respectively); and to $a=10 {\rm \AA}$,
with $\Delta k = 0.02 {\rm \AA}^{-1}$
and 0.04 ${\rm \AA}^{-1}$ (plots 3 and 4, respectively), the mean
kinetic energy $\overline E$ being 5 eV, i.e., one half of $V_{0}$. \ \
In Fig.5 the plots are shown
of $<\tau_{\, \rm Ret}(x,x)>$. \ The curves 1, 2 and 3 correspond to
$\overline E = 2.5$ eV, $\;$5 eV and 7.5 eV, respectively, for $\Delta k =
0.02 {\rm \AA}^{-1}$ and  $a = 5  {\rm \AA}$; \ the curves 4, 5 and 6
correspond to $\overline E = 2.5$ eV, $\;$5 eV  and 7.5 eV, respectively, for
$\Delta k = 0.04 {\rm \AA}^{-1}$ and $a = 5 {\rm \AA}$; \ while the curves
7, 8 and 9
correspond to $\Delta k =0.02 {\rm \AA}^{-1}$ and 0.04 ${\rm \AA}^{-1}$,
respectively, for $\overline E = 5$ eV and $a=10 {\rm \AA}$.

\h Also from the new Figs.3--5 one can see that: \ 1) at variance with
ref.[8], no plot considered by us for the mean penetration duration
$<\tau_{\, \rm Pen}(0,x)>$ of our wave packets presents  any interval
with negative values,
nor with a decreasing  $<\tau_{\, \rm Pen}(0,x)>$
for increasing  $x$; \ and, moreover, that \  2) the mean tunnelling
duration
$<\tau_{\rm Tun}(0,a)>$  does not depend on the barrier  width $a$
(``Hartman effect"); \ and finally that \ 3) quantity
$<\tau_{\rm Tun}(0,a)>$ decreases when the energy increases. \ \ Furthermore,
it is noticeable that also from Figs.3--5 we observe: \ 4) a rapid increase
for the value of the electron penetration time in the
initial part of the barrier region (near $x= 0$); \ and \ 5) a tendency of
$<\tau_{\, \rm Pen}(0,x)>$ to a saturation value in the final part of
the barrier, near $x=a$.

\h Feature 2), firstly
observed for quasi-monochromatic particles,$^{[2]}$
does evidently agree with the predictions made in ref.[1] for
arbitrary wave packets. \ Feature 3) is also in agreement with previous
evaluations performed
for quasi-monochromatic particles and presented, for instance, in
refs.[1,2,15]. \ Features 4) and 5) can be apparently explained by
interference between
those initial penetrating and returning waves inside
the barrier, whose superposition yields the resulting fluxes $J_{+}$ and
$J_{-}$. In particular, if in the initial part of the barrier
the returning--wave packet is comparatively large, it
does essentially extinguish the leading edge
of the incoming--wave packet. \ By contrast, if for growing $x$
the returning--wave packet quickly vanishes, then the contribution
of the leading edge of the incoming--wave packet to the mean penetration
duration $<\tau_{\, \rm Pen} (0,x)>$
does initially (quickly) grow, while in the final barrier region its increase
does rapidly slow down.

\h Furthermore, the larger is the barrier width $a$, the larger is the part of
the
back edge of the incoming--wave packet which is
extinguished by interference with the returning--wave packet. \
Quantitatively,
these phenomena will be studied elsewhere. \ Finally, in connection with
the plots
of $<\tau_{\, \rm Ret}(x,x)>$ as a function of $x$, presented in Fig.5,
let us observe that: \ (i) the mean reflection duration \
$<\tau_{\, \rm R} (0,0)> \; \equiv \; <\tau_{\, \rm Ret} (0,0)>$ \
does not depend on the barrier width $a$;  \ (ii) in correspondence
with the barrier region betwen $0$ and approximately $0.6 \ a$,
the value of $<\tau_{\, \rm Ret}(0,x)>$ is almost constant; \ while \ (iii)
its value increases with $x$ only in the barrier region near $x = a$ \
(even if
it should be pointed out that our calculations near $x = a$  are not so good,
due to the very small values assumed by \
$\int_{-\infty}^{\infty} J_{-}(x,t) \drm t$ \ therein). \ \ Let us notice that
point (i), also
observed firstly for quasi-monochromatic particles,$^{[2]}$
is as well in accordance with the results obtained in ref.[1] for arbitrary
wave packets. \ Moreover, also points (ii) and (iii) can be explained by
interference phenomena inside the barrier: if, near $x=a$, the initial
returning--wave packet is almost totally quenched by the initial
incoming--wave packet, then only a negligibly small piece of its
back edge (consisting of the components with the smallest velocities) does
remain. \ With decreasing $x$
\ ($x \rightarrow 0$), \ the unquenched part of the returning--wave packet
seems to become more and more large (containing more and more rapid
components),
thus making the difference \ $<\tau_{\, \rm Ret}(0,x)> -
<\tau_{\, \rm Pen}(0,x)>$ \ almost constant. \
And the interference  between incoming and reflected waves at points
$x \le 0$ does effectively constitute a retarding phenomenon \ [so that \
$<t_-(x=0)>$ is larger than $<\tau_{\, \rm R}(x=0)$], \ which can explain
the larger values of $<\tau_{\, \rm R}(x=0,x=0)>$ in comparison    with
$<\tau_{\rm Tun}(x=0,x=a)>$.

\h Therefore our evaluations, in all the cases considered above,
appear to confirm our previous analysis
at page 352 of ref.[1], and our conclusions therein concerning in particular
the validity of the Hartman effect also for {\em non}--quasi-monochromatic
wave packets. \ Even more, since the interference between incoming and
reflected waves before the barrier (or between penetrating and returning
waves, inside the barrier, near the entrance wall) does just {\em increase}
the tunnelling time as well as the transmission times,
we can expect that our non-relativistic formulae for
$<\tau_{\rm Tun}(0,a)>$  and  $<\tau_{\rm T}(x_{\rm i} < 0, x_{\rm f} > a)>$
will always forward positive values.$^{\# 3}$\\
\footnotetext{$^{\# 3}$ A different claim by Delgado, Brouard and
Muga$^{[17]}$ does not seem to be relevant to our calculations, since it is
based once more, like ref.[8],  not on our but on different wave packets (and
over--barier components are also retained in ref.[17], at variance
with us). \  Moreover, in their classical example, they overlook the fact that
the mean entrance time $<t_+(0)>$ gets contribution  mainly by the rapid
components of the wave packet; they forget, in fact, that the slow components
are (almost) totally reflected by the
initial wall, causing a quantum--mechanical reshaping that contributes to the
initial ``time decrease" discussed by us already in the last few paragraphs
of page 352 in ref.[1]. \ All such phenomena {\em reduce} the value of
$<t_+(0)>$, and we expect it to be (in our non-relativistic treatment)
less than  $<t_+(a)>$.}

\h At this point, it is necessary ---however--- to observe the following.
 \ Even if our non-relativistic equations are not expected (as we have
just seen)  to yield negative times, nevertheless
one ought to bear in mind that (whenever
it is met an object, ${\cal O}$, travelling at Superluminal speed) negative
contributions should be expected
to the tunnelling times: and this ought not to be
regarded as unphysical.  In fact, whenever an ``object" ${\cal O}$ {\em
overcomes}
the infinite speed$^{[18]}$ with respect to a certain observer,
it will afterwards appear to the same observer as its ``{\em anti}-object"
$\ove{\cal O}$ travelling in the opposite {\em space} direction$^{[18]}$.  \
For instance, when passing from the lab to a frame ${\cal F}$ moving in
the {\em same}
direction as the particles or waves entering the barrier region, the
objects $\cal O$ penetrating through the final part of the barrier (with
almost infinite speeds, like in Figs.1--5)
will appear in the frame ${\cal F}$ as anti-objects $\ove{\cal O}$
crossing that portion of the barrier {\em in the opposite
space--direction}$^{[18]}$.  In the new frame ${\cal F}$,
therefore, such anti-objects $\ove {\cal O}$ would yield a
{\em negative} contribution to the
tunnelling time: which could even result, in total,  to be negative. \
For any clarifications, see refs.[18]. \ So, we have no
objections a priori against the fact that Leavens can find, in certain cases,
negative values$^{[8,17]}$:
e.g, when applying our formulae to wave packets with suitable
initial conditions. \ What we want to stress here is that the appearance
of negative times (it being predicted by Relativity itself,$^{[18]}$  when
in presence
of {\em anything} travelling faster than $c$) is not a valid reason to rule
out a theoretical approach.\\

\h {\em At last}, let us ---incidentally--- recall and mention the following
fact. Some preliminary
calculations of penetration times (inside a rectangular barrier)
for tunnelling gaussian wave packets had been presented by us in 1994
in ref.[16].  Later on ---looking for
any possible explanations for the disagreement between the results in
ref.[8] and in our ref.[16]--- we discovered,
however, that an exponential factor was missing in a term of one of the
fundamental formulae on which
the numerical computations (performed by our group in Kiev) were based: a
mistake that
could not be detected, of course, by our careful checks about the computing
process. \ \ Therefore, the new results of ours appearing
in Figs.1--2 should {\em replace} Figs.1--3 of ref.[16].
\ One may observe that, by using the same parameters as (or
parameters very
near to) the ones adopted by Leavens for his Figs.3 and 4 in ref.[8],
our new, corrected figures 1 and 2 result to be more similar to Leavens'
than the uncorrected ones (and this is of course a welcome step towards
the solution of the problem). \ One can verify once
more, however, that
our theory appears to yield {\em for those parameters} non-negative results
for $<\tau_{\, \rm Pen} (x_{\rm f})>$, contrarily to a claim in ref.[8].
 \ Actually, our previous general
conclusions have not been apparently affected by the mentioned mistake. \ In
particular,
the value of $<\tau_{\, \rm Pen} (x_{\rm f})>$ increases with increasing
$x_{\rm f}$, and tends to saturation for $x_{\rm f} \rightarrow a$. \
We acknowledge, however, that the difference in the adopted integration
ranges
[$- \infty$ to $+ \infty$ for us, and $0$ to $+ \infty$ for Leavens] does
not play an important role, contrarily to our previous belief,$^{[16]}$ in
explaining the remaining discrepancy between our results and Leavens'. Such
a discrepancy {\em might} perhaps  depend on the fact
that the functions to be integrated do fluctuate heavily$^{\# 4}$ (anyway,
\footnotetext{$^{\# 4}$  We can {\em only}
say that we succeeded in reproducing  results of the type put forth in
ref.[8] by using larger steps; whilst the {\em ``non-causal"} results
disappeared ---{\em in the considered cases}--- when adopting small enough
integration steps.}
we did carefully check  that our own
elementary integration step in the integration over $\drm k$
was small enough  in order to
guarantee the stability of the numerical result, and, in particular, of
their sign, for strongly oscillating functions in the integrand). \ More
probably, the persisting disagreement
can be merely due to the fact ---as recently claimed also
by Delgado et al.$^{[17]}$--- that {\em different} initial conditions
for the wave packets
were actually chosen in ref.[8] and in ref.[1].  \ Anyway,
our approach seems to get support, at least in some particular cases,
also by a recent article by Brouard et al., which ``generalizes" ---even if
starting from a totally different point of view--- some of our
results.$^{[12,15]}$.

\h Let us take advantage of the present opportunity for answering other
criticisms appeared in
ref.[8], where it has been furthermore commented about our way of
performing actual averages over the physical time.  We cannot agree with
those comments: let us re-emphasize in fact that, within conventional quantum
mechanics, the time $t(x)$ at which our particle (wave packet) passes through
the position $x$ is ``statistically distributed" with the probability
densities \ ${\rm d}t J_{\pm}(x,t) / \int_{-\infty}^{\infty} {\rm d}t
J_{\pm}(x,t)$, \ as we explained at page 350 of ref.[1]. \ This
distribution meets the requirements of the time--energy uncertainty
relation.

\h We also answered in Sect.{\bf 1} Leavens' comments about
our analysis$^{[1]}$ of the dwell--time approaches.$^{[19]}$

\h The last object of the criticism in ref.[8] refers to the
impossibility,
in our approach, of distinguishing between ``to be transmitted" and ``to be
reflected" wave packets at the leading edge of the barrier. \ Actually, we do
{\em distinguish} between them; only, we cannot ---of course---
{\em separate} them,
due to the obvious {\em superposition} (and interference) of both
wave functions in
$\rho (x,t)$, in $J(x,t)$ and even in $J_{\pm} (x,t)$.  \ This is known to be
an unavoidable consequence of the superposition principle, valid for {\em wave
functions} in conventional quantum mechanics. \ That last objection, therefore,
should be addressed to quantum mechanics, rather then to us. \ \
Nevertheless, Leavens' aim of comparing the definitions proposed
by us for the tunnelling times not only with conventional, but also with
non--standard quantum mechanics {\em might} be regarded a priori as
stimulating and possibly worth of further investigation.

\vspace*{1 cm}

{ \bf 4. -- Further remarks}

\vspace*{0.5 cm}

\h In connection with the question of ``causality" for relativistic
tunnelling particles, let us stress that
the Hartman-Fletcher phenomenon (very small tunnelling
durations), with the consequence of Superluminal velocities for
sufficiently wide barriers, was found
theoretically also in QFT for Klein--Gordon
and Dirac equations,$^{[1]}$ and experimentally for electromagnetic
evanescent--mode wave packets$^{[4-6]}$ (tunnelling photons). \
It should be recalled that the problem of
Superluminal velocities for electromagnetic wave packets in media
with anomalous dispersion, with absorption, or behaving as a barrier
for photons (such as regions with frustrated internal reflection)
has been present in the scientific literature since long (see, for
instance, quotations [2,1,18], and refs. therein); even if
a complete settlement of the causal problem is not yet available
for relativistic waves (differently from the case
of point particles$^{[18]}$). \
Apparently, it is not sufficient to pay attention only to
group velocity and mean duration for a particle passing through
a medium; on the contrary, it is important taking into
account and studying ab initio the {\em variances} (and the higher order
central moments) of the duration distributions,
as well as the wave packet {\em reshaping} in presence of a barrier, or
inside anomalous media (even if reshaping does {\em not} play {\em always} an
essential role).

\h Passing to the approaches {\em alternative} to the
direct description of tunnelling processes in terms of
wave packets, let us here recall those ones
which are based  on averaging over
the set of all dynamical paths (through the Feynman path integral
formulation, the
Wigner distribution method, and the non-conventional Bohm approach), and
others
that use additional degrees of freedom  which can be used as ``clocks".
General analyses of all such alternative approaches can be found in
refs.[1, 20--24] from different points of view.

\h If one confines himself within the
framework of  conventional quantum mechanics, then the Feynman path
integral formulation seems to be adequate.$^{[24]}$ \ But it is not clear
what procedure is
needed to calculate physical quantities within the Feynman--type
approach$^{[23]}$,
and usually such calculations result in complex tunnelling durations.
The Feynman approach seems to need further modifications if one wants
to apply it to the time analysis of tunnelling processes, and its results
obtaind up to now cannot be considered as final.

\h As to the approaches based on introducing additional
degrees of freedom
as ``clocks", one can often realize that the tunnelling time happens to be
noticeably distorted by the presence of such degrees of freedom. For example,
the B\"uttiker-Landauer time is connected with absorption or emission
of modulation quanta (caused by the time--dependent oscillating part
of the barrier potential) during tunnelling, rather than with the
tunnelling process itself.$^{[1,15]}$  \ And, with reference to the
Larmor precession time, it has been
shown$^{[11,20]}$ that this time definition is connected not
only with the intrinsic tunnelling process, but also with the geometric
boundaries
of the magnetic field introduced as a part of the clock: for instance, if
the magnetic field region is
infinite, one ends up with the phase tunnelling time, after an average
over the (small) energy spread of the wave packet.  \ Actually, those
``clock" approaches, when applied to tunnelling wave packets, seem to lead
---after eliminating the distortion caused by the
additional degrees  of freedom--- to the same results
as the direct wave packet approach, whatever be the weight function adopted
in the time integration.

\vspace*{1 cm}

{\bf Acknowledgements:} The authors thank M. Pignanelli, G.M. Prosperi,
G. Salesi,
V.S. Sergeyev, S. Sambataro, M.T. Vasconselos, A. Vitale,
and B.N. Zakhariev for their scientific
collaboration; and A. Agresti, M. Baldo, E. Beltrametti, A. Bugini,
G. Giardina, G. Giuffrida,
L. Lo Monaco, V.L. Lyuboshitz, G.D. Maccarrone, R.L. Monaco,
J.G. Muga, G. Nimtz, T.V. Obikhod,
E.C. Oliveira, E. Parigi, R. Pucci, A. Ranfagni, W.A. Rodrigues,
M. Sambataro,  P. Saurgnani, J. Vaz for useful discussions or cooperation. \
At last, they
thank the Editor for valuable comments, and J.G. Muga for having sent them
a preprint of his before publication.

\newpage

\centerline{{\bf Figure Captions}}

\vspace*{1. cm}

{\bf Fig.1} -- Behaviour of the average penetration time
$<\tau_{\, \rm Pen}(0,x)>$
(expressed in seconds) as a function of the penetration depth
$x_{\rm f} \equiv x$
(expressed in {\aa}ngstroms) through a rectangular
barrier with width $a = 5 \; {\rm \AA}$, for \
$\Delta k = 0.02 \; {\rm {\AA}}^{-1}$ (dashed line)
 \ and \ $\Delta k = 0.01 \; {\rm {\AA}}^{-1}$ (continuous line),
\ respectively. \ The other parameters are listed in footnote $\# 1$.
 \ It is worthwhile to notice that $<\tau_{\, \rm Pen}>$ rapidly increases for
the first, few initial {\aa}ngstroms ($\sim 2.5$ \AA), tending afterwards to a
saturation value. This seems to confirm the existence of the so-called
``Hartman effect".$^{[2,1,15]}$

\vspace*{0.6 cm}

{\bf Fig.2} -- The same plot as in Fig.1, for \ $\Delta k = 0.01 \;
{\rm {\AA}}^{-1}$, \  except that now the barrier width
is $a = 10 \; {\rm \AA}$. \ Let us observe that the numerical values of the
(total) tunnelling time
$<\tau_{\rm T}>$ practically does not change when passing from \
$a = 5 \;$\AA  \ to \ $a = 10 \;$\AA, \ again in agreement with the
characteristic features$^{1}$
of the Hartman effect. \ Figures 1 and 2 do improve (and correct) the
corresponding ones, preliminarily presented by us in ref.[16].

\vspace*{0.6 cm}

{\bf Fig.3} -- Behaviour of $<\tau_{\, \rm Pen}(0,x)>$ (expressed in seconds)
as a
function of $x$ (expressed in {\aa}ngstroms), for tunnelling through a
rectangular barrier with
width $a = 5 \;${\AA} and for different values of ${\ove E}$ and of
$\Delta k$:

curve 1: $\Delta k = 0.02 {\rm \AA}^{-1}$ and $\ove E = 2.5 \;
{\rm eV}$; \ \
curve 2: $\Delta k = 0.02 {\rm \AA}^{-1}$ and $\ove E = 5.0 \;
{\rm eV}$; \ \
curve 3: $\Delta k = 0.02 {\rm \AA}^{-1}$ and $\ove E = 7.5 \;
{\rm eV}$; \ \
curve 4: $\Delta k = 0.04 {\rm \AA}^{-1}$ and $\ove E = 5.0 \;
{\rm eV}$.

\vspace*{0.6 cm}

{\bf Fig.4} -- Behaviour of $<\tau_{\, \rm Pen}(0,x)>$ (in seconds) as a
function of $x$ (in {\aa}ngstroms)  for $\ove E = 5 \; {\rm eV}$
and different values of $a$ and $\Delta k$:

curve 1: $a = 5  \; {\rm \AA}$ and $\Delta k = 0.02 {\rm \AA}^{-1}$; \ \
curve 2: $a = 5  \; {\rm \AA}$ and $\Delta k = 0.04 {\rm \AA}^{-1}$; \ \
curve 3: $a = 10 \; {\rm \AA}$ and $\Delta k = 0.02 {\rm \AA}^{-1}$; \ \
curve 4: $a = 10 \; {\rm \AA}$ and $\Delta k = 0.04 {\rm \AA}^{-1}$.

\vspace*{0.6 cm}

{\bf Fig.5} --Behaviour of $<\tau_{\, \rm Ret}(x,x)>$ (in seconds) as a
function of $x$ (in {\aa}ngstroms)  for
different values of $a$,  $\ove E$ and $\Delta k$:

curve 1: $a = 5 \; {\rm \AA}$, \ $\ove E = 2.5 \; {\rm eV}$ and
$\Delta k = 0.02 {\rm \AA}^{-1}$; \ \
curve 2: $a = 5 \; {\rm \AA}$, \ $\ove E = 5.0 \; {\rm eV}$ and
$\Delta k = 0.02 {\rm \AA}^{-1}$; \ \
curve 3: $a = 5 \; {\rm \AA}$, \ $\ove E = 7.5 \; {\rm eV}$ and
$\Delta k = 0.02 {\rm \AA}^{-1}$; \ \
curve 4: $a = 5 \; {\rm \AA}$, \ $\ove E = 2.5 \; {\rm eV}$ and
$\Delta k = 0.02 {\rm \AA}^{-1}$; \ \
curve 5: $a = 5 \; {\rm \AA}$, \ $\ove E = 5.0 \; {\rm eV}$ and
$\Delta k = 0.02 {\rm \AA}^{-1}$; \ \
curve 6: $a = 5 \; {\rm \AA}$, \ $\ove E = 7.5 \; {\rm eV}$ and
$\Delta k = 0.02 {\rm \AA}^{-1}$; \ \
curve 7: $a = 5 \; {\rm \AA}$, \ $\ove E = 5.0 \; {\rm eV}$ and
$\Delta k = 0.02 {\rm \AA}^{-1}$; \ \
curve 8: $a = 5 \; {\rm \AA}$, \ $\ove E = 5.0 \; {\rm eV}$ and
$\Delta k = 0.02 {\rm \AA}^{-1}$.


\newpage

\begin{thebibliography}{24}

\bibitem{1} V.S. Olkhovsky and E. Recami: Physics Reports {\bf 214}
(1992) 339.

\bibitem{2} T.E. Hartman: J. Appl. Phys. {\bf 33} (1962) 3427; \ J.R.
Fletcher: J. Phys. C{\bf 18} (1985) L55. \ See also C.G.B. Garret and D.E.
McCumber: Phys. Rev. A{\bf 1} (1970) 305; \ S. Chu and S. Wong: Phys. Rev.
Lett.
{\bf 48} (1982) 738; \ S. Bosanac: Phys. Rev. A{\bf 28} (1983) 577; \
F.E. Low and P.F. Mende: Ann. of Phys. {\bf 210} (1991) 380.

\bibitem{3} See, e.g., Th. Martin and R. Landauer: Phys. Rev. A{\bf 45}
(1992) 2611; \ R.Y. Chiao, P.G. Kwiat and A.M. Steinberg: Physica B{\bf 175}
(1991) 257; \ A. Ranfagni, D. Mugnai, P. Fabeni and
G.P. Pazzi: Appl. Phys. Lett. {\bf 58} (1991) 774.

\bibitem{4} A. Enders and G. Nimtz: J. Physique I {\bf 2} (1992) 1693; \
{\bf 3} (1993) 1089; \ Phys. Rev. B{\bf 47} (1993) 9605;
 \ E{\bf 48} (1993) 632; \ J. Physique I {\bf 4} (1994) 1817; \
G. Nimtz, A. Enders and H. Spieker: J. Physique I {\bf 4}
(1994) 1; \ ``Photonic tunnelling experiments: Superluminal tunnelling", in
{\em Wave and particle in light and matter (Proceedings of the Trani
Workshop, Italy, Sept.$\!$ 1992)}, ed. by A. van der Merwe and A. Garuccio
(Plenum; New York, in press); \ W. Heitmann and G. Nimtz: Phys. Lett.
A{\bf 196} (1994) 154.

\bibitem{5} A.M. Steinberg, P.G. Kwiat and R.Y. Chiao: Phys. Rev. Lett.
{\bf 71} (1993) 708, and refs. therein; \ Scientific American {\bf 269} (1993)
issue no.2, p.38. \ See also P.G. Kwiat, A.M. Steinberg, R.Y.Chiao, P.H.
Eberhard and M.D. Petroff: Phys. Rev. A{\bf 48} (1993) R867; \ E.L. Bolda,
R.Y. Chiao and J.C. Garrison: Phys. Rev. A{\bf 48} (1993) 3890.

\bi{6} A. Ranfagni, P. Fabeni, G.P. Pazzi and D. Mugnani: Phys. Rev.
E{\bf 48} (1993) 1453; \ Ch. Spielmann, R. Szip\"ocs, A. Stingl and F.
Krausz: Phys. Rev. Lett. {\bf 73} (1994) 2308. \ Cf. also J. Brown: New
Scientist (April, 1995), p.26.

\bi{7} W. Jaworski and D.M. Wardlaw: Phys. Rev. A{\bf 37} (1988) 2843.

\bibitem{8} C.R. Leavens: Solid State Commun. {\bf 85} (1993) 115.

\bibitem{9} R. Landauer and Th. Martin: Solid State Commun. {\bf 84}
(1992) 115.

\bibitem{10} R.S. Dumont and T.L. Marchioro: Phys. Rev. A{\bf 47} (1993) 85.

\bibitem{11} See {\em e.g.} V.S. Olkhovsky: Nukleonika {\bf 35} (1990) 99, and
refs. therein; in particular, V.S. Olkhovsky: Doctorate (Habilitation) Thesis,
Institute for Nuclear Research, Ukrainian Academy of Sciences, Kiev (1986). \
See also V.S. Olkhovsky, V.M. Shilov and B.N. Zakhariev: Oper. Theory Adv.
Appl. {\bf 46} (1990) 159.

\bibitem{12} S.Brouard, R. Sala and J.G. Muga: Phys.Rev. A{\bf 49} (1994)
4312; \  some criticism to this paper appeared in  C.R. Leavens: Phys.
Lett. A{\bf 197} (1995) 88. \ Cf. also A.F.M. Anwar and M.M. Jahan: IEEE J.
Quantum Electronics {\bf 31} (1995) 3.

\bibitem{13} L. Landau and E.M. Lifshitz: {\em Quantum Mechanics},
3rd ed. (Pergamon Press; Oxford, 1977), Sect.{\bf 20}.

\bibitem{14} C.R. Leavens: Phys. Lett. A{\bf 197} (1995) 88.

\bibitem{15} V.S. Olkhovsky, E. Recami and A.K. Zaichenko: Report
INFN/FM--94/01 (Frascati, 1994). \ See also F. Raciti and G. Salesi:
J. de Phys. I {\bf 4} (1994) 1783.

\bibitem{16}  V.S. Olkhovsky, E.Recami and A.K. Zaichenko:
Solid State Commun. {\bf 89} (1994) 31.

\bibitem{17} V. Delgado, S. Brouard and J.G. Muga: ``Does positive flux
provide a valid definition of tunnelling time?", to appear in
Solid State Commun.

\bibitem{18} E. Recami: ``Classical tachyons and possible applications",
Rivista Nuovo Cim. {\bf 9} (1986), issue no.6, pp.1-178; \ E. Recami:
``A systematic, thorough analysis of the tachyon causal paradoxes",
Found. of Phys. {\bf 17} (1987)  239-296; \ E.Recami: ``The Tolman--Regge
{\em antitelephone} paradox: Its solution by tachyon dynamics", Lett. Nuovo
Cim. {\bf 44} (1985) 587.

\bibitem{19} F. T. Smith: Phys. Rev. {\bf 118} (1960) 349; \
M. Buttiker: Phys. Rev. B{\bf 27} (1983) 6178; \
M.L. Goldberger and K.M. Watson: {\em Collision Theory} (Wiley;
New York, 1964); \
J.M. Jauch and J.P. Marchand: Helv. Phys. Acta {\bf 40}
(1967) 217; \
Ph.A. Martin: Acta Phys. Austr. Suppl. {\bf 23} (1981) 157.

\bibitem{20}  Z.H. Huang, P.H. Cutler, T.E. Feuchtwang, E. Kazes, H.Q. Nguen
and T.E. Sullivan, J. Vac. Sci. Technol. A{\bf 8} (1990) 186.

\bibitem{21} C.R. Leavens and G.C. Aers: in {\em Scanning Tunnelling
Microscopy and Related Methods}, edited by  R.J. Behm, N. Garcia and
H. Rohrer (Kluwer; Dordrecht, 1990), p.59.

\bibitem{22} A.P. Jauho: in {\em Hot Carriers in Semiconductor nanostructures,
Physics and Applications}, edited by J. Shah ( Academic Press; Boston, 1992),
p.121.

\bibitem{23} R. Landauer and Th. Martin: Rev. Mod. Phys. {\bf 66} (1994) 217.
 \ Cf. also E.H. Hauge and J.A. Stvneng: Rev. Mod. Phys. {\bf 71}
(1989) 917.

\bibitem{24} W. Jaworski and D.M. Wardlaw: Phys. Rev. A{\bf 48} (1993) 3375.

\end{thebibliography}
\end{document}